\documentclass[aps,manuscript,pop,showpacs,preprint,superscriptaddress]{revtex4-1}
\usepackage{lmodern}

\usepackage[T1]{fontenc}
\usepackage[latin9]{inputenc}
\usepackage{amsmath}
\usepackage{graphicx}

\makeatletter
\usepackage{amsthm}\usepackage{latexsym}\usepackage{bm}\usepackage{amsfonts}

\makeatother

\begin{document}

\title{On the physical mechanism of three-wave instabilities\textendash{}
resonance between positive- and negative-action modes}

\author{Ruili Zhang}

\affiliation{School of Science, Beijing Jiaotong University, Beijing 100044, China}

\author{Hong Qin }

\thanks{Corresponding author, hongqin@princeton.edu}

\affiliation{Plasma Physics Laboratory, Princeton University, Princeton, NJ 08543,
USA}

\affiliation{Department of Modern Physics, University of Science and Technology
of China, Hefei, Anhui 230026, China}

\author{Yuan Shi }

\affiliation{Plasma Physics Laboratory, Princeton University, Princeton, NJ 08543,
USA}

\author{Jian Liu}

\affiliation{Department of Modern Physics, University of Science and Technology
of China, Hefei, Anhui 230026, China}

\author{Jianyuan Xiao}

\affiliation{Department of Modern Physics, University of Science and Technology
of China, Hefei, Anhui 230026, China}
\begin{abstract}
Three-wave instability is a fundamental process that has important
applications in many branches of physics. It is widely accepted that
the resonant condition $\omega_{z}\approx\omega_{x}+\omega_{y}$ for
participating waves is the criteria for the onset of the instability.
We show that this condition is neither sufficient nor necessary, instead,
the exact criteria for the onset of the instability is that a positive-action
mode resonates with a negative-action mode. This mechanism is imposed
by the topology and geometry of the spectral space. Guided by this
new theory, additional instability bands previously unknown are discovered.
\end{abstract}
\maketitle
Three-wave interaction is a fundamental nonlinear process in complex
media that has important applications in different branches of physics.
In plasma physics, three-wave interaction and the associated parametric
decay instability have been systematically studied for both magnetized
and unmagnetized plasmas \cite{oraevskii1963stability,silin1965parametric,nishikawa1968parametric,nishikawa1968parametric2,rosenbluth1973temporal,drake1974parametric,kaup1979space,jha1975temporal,malkin1999fast,dodin2002storing,hasegawa2012plasma,kaw2017nonlinear}.
Since first proposed in 1990s \cite{malkin1999fast}, it has been
successfully applied to compress laser pulses \cite{ping2004amplification,shi2017laser}.
In optics, three-wave interaction was identified in stimulated Raman
and Brillouin scattering \cite{shen1965theory,bloembergen1964coupling,blow1989theoretical,singh2007nonlinear}
and more recently in second harmonic generation \cite{leo2016walk},
optical soliton formation \cite{picozzi2001parametric,buryak2002optical},
and optical rouge waves \cite{chen2015optical,chen2015watch}. Three-wave
interaction plays an important role in fluid dynamics as well \cite{craik1988wave}.
For example, it has been studied experimentally for gravity-capillary
waves in recent years \cite{aubourg2015nonlocal,haudin2016experimental,bonnefoy2016observation}.

For the three-wave interaction to be unstable, it is generally believed
that the frequencies of the participating waves $\omega_{x}$, $\omega_{y}$,
and $\omega_{z}$ need to satisfy the resonant condition \cite{nishikawa1968parametric,nishikawa1968parametric2,drake1974parametric}
\begin{equation}
\omega_{z}\approx\omega_{x}+\omega_{y}.\label{eq:rc}
\end{equation}
In this paper, we show that in general condition \eqref{eq:rc} is
neither sufficient nor necessary for the onset of the instability.
Instead, the three-wave instability is triggered when and only when
a positive-action mode resonates with a negative-action mode, and
the frequencies of the positive- and negative-action modes are in
general different from those of the participating waves. Only in the
weak interaction limit, the familiar resonant condition \eqref{eq:rc}
is recovered within one of the unstable regions.

\begin{figure}
\includegraphics[scale=0.65]{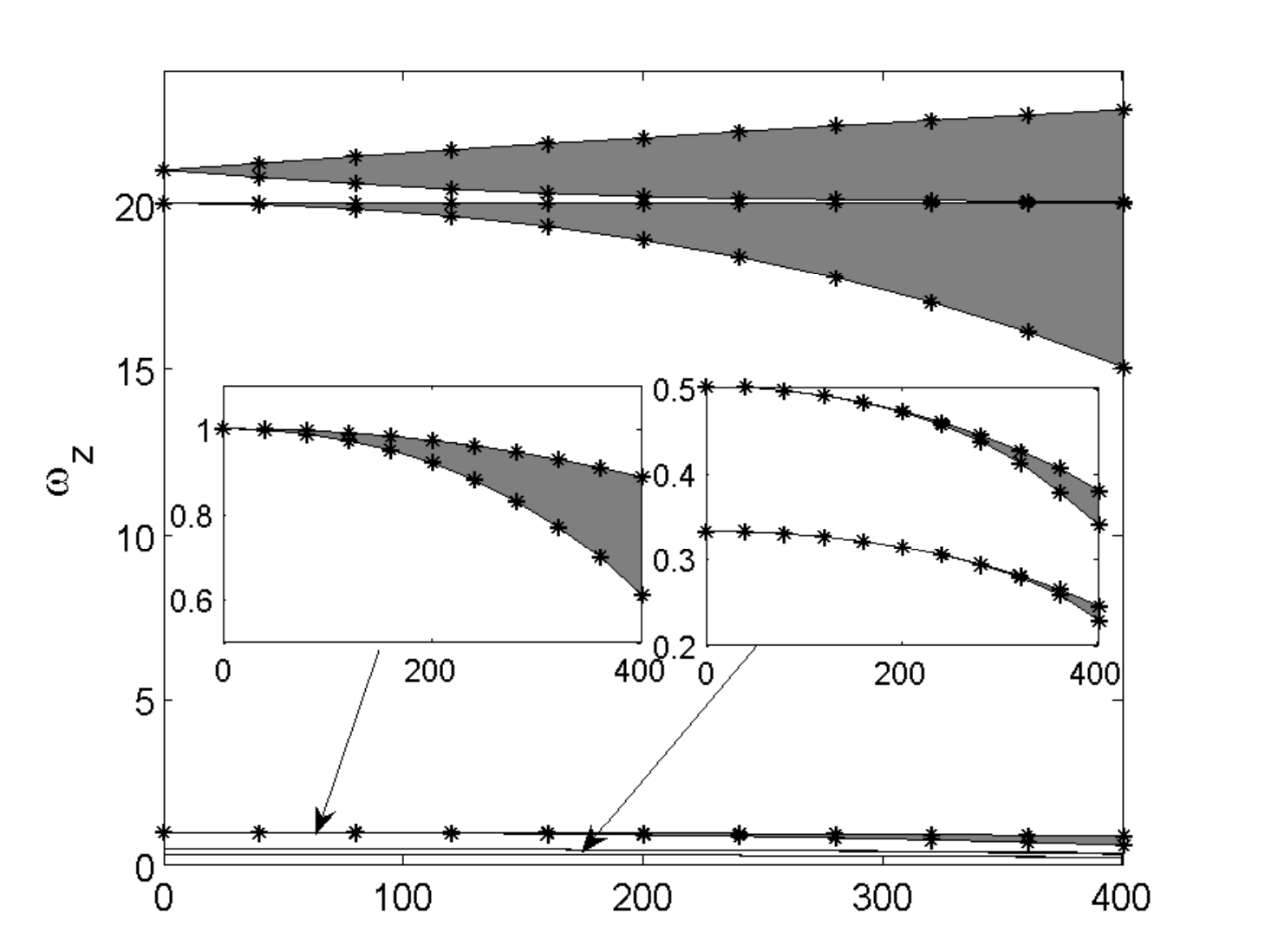}

\caption{Plot of instability regions represented by the normalized frequency
of the pump wave $\omega_{z}$ as a function of the normalized strength
of the pump wave $E_{0}$. Parameters are chosen for the three-wave
interaction in a typical hydrogen plasma with $\omega_{x}=1$, $\omega_{y}=20$,
$\lambda_{x}=i\mu_{x}$ and $\lambda_{y}=i\mu_{y}$ with $\mu_{x}=-\mu_{y}m_{e}/m_{i}=-\mu_{y}/1836$
and $\mu_{y}=E_{0}.$ Shaded regions are unstable. Five instability
bands are shown. There exit numerous narrow instability bands under
the lowest band shown. The top three bands originate from $\omega_{z}=\omega_{x}+\omega_{y}$,
$\omega_{z}=\omega_{y}$ and $\omega_{z}=\omega_{x}$ on the vertical
axis. Only the uppermost instability band originates from the position
satisfying condition \eqref{eq:rc} on the vertical axis. }
\label{f0}
\end{figure}

\begin{figure}
\includegraphics[scale=0.6]{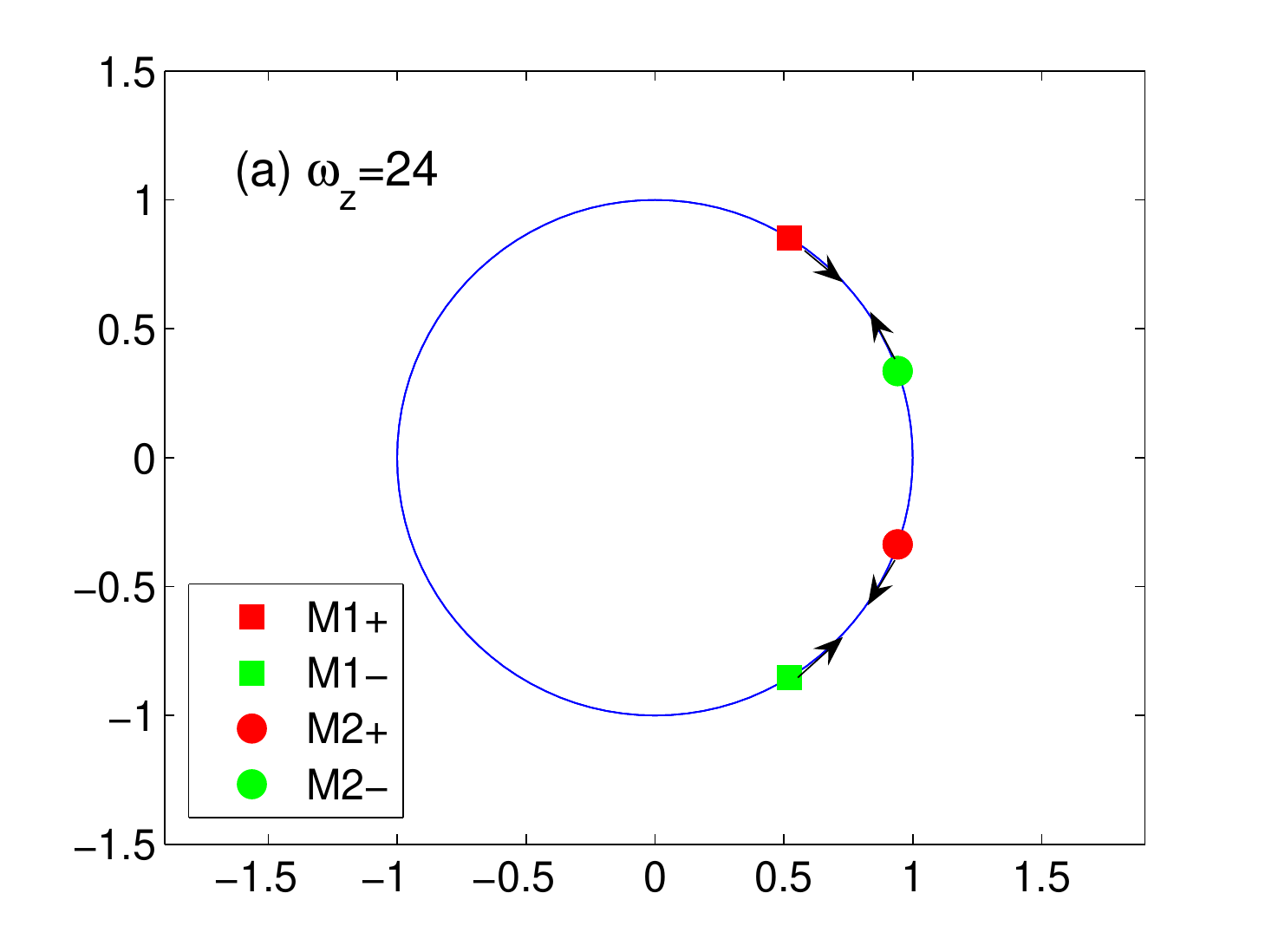}\includegraphics[scale=0.6]{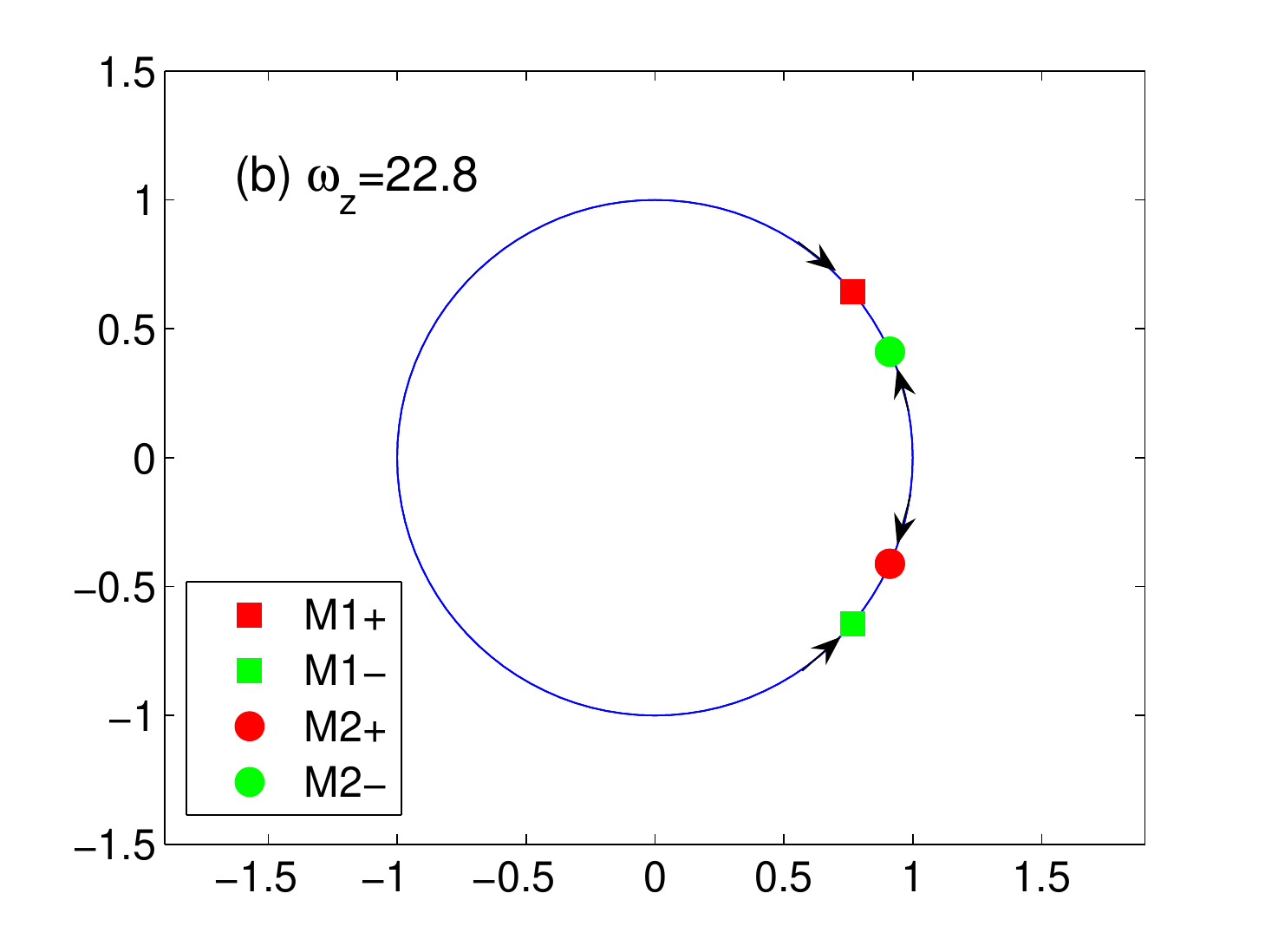}

\includegraphics[scale=0.6]{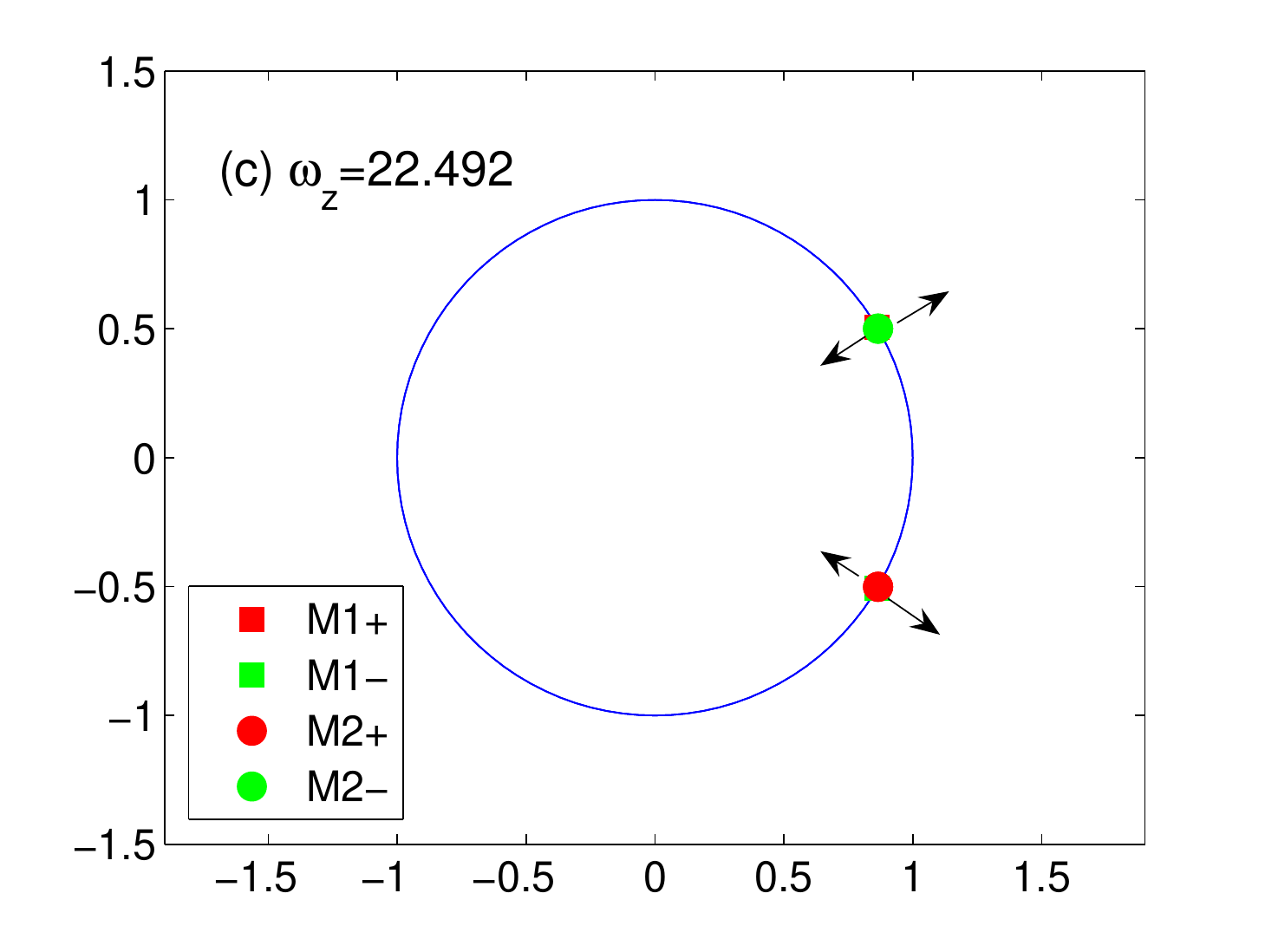}\includegraphics[scale=0.6]{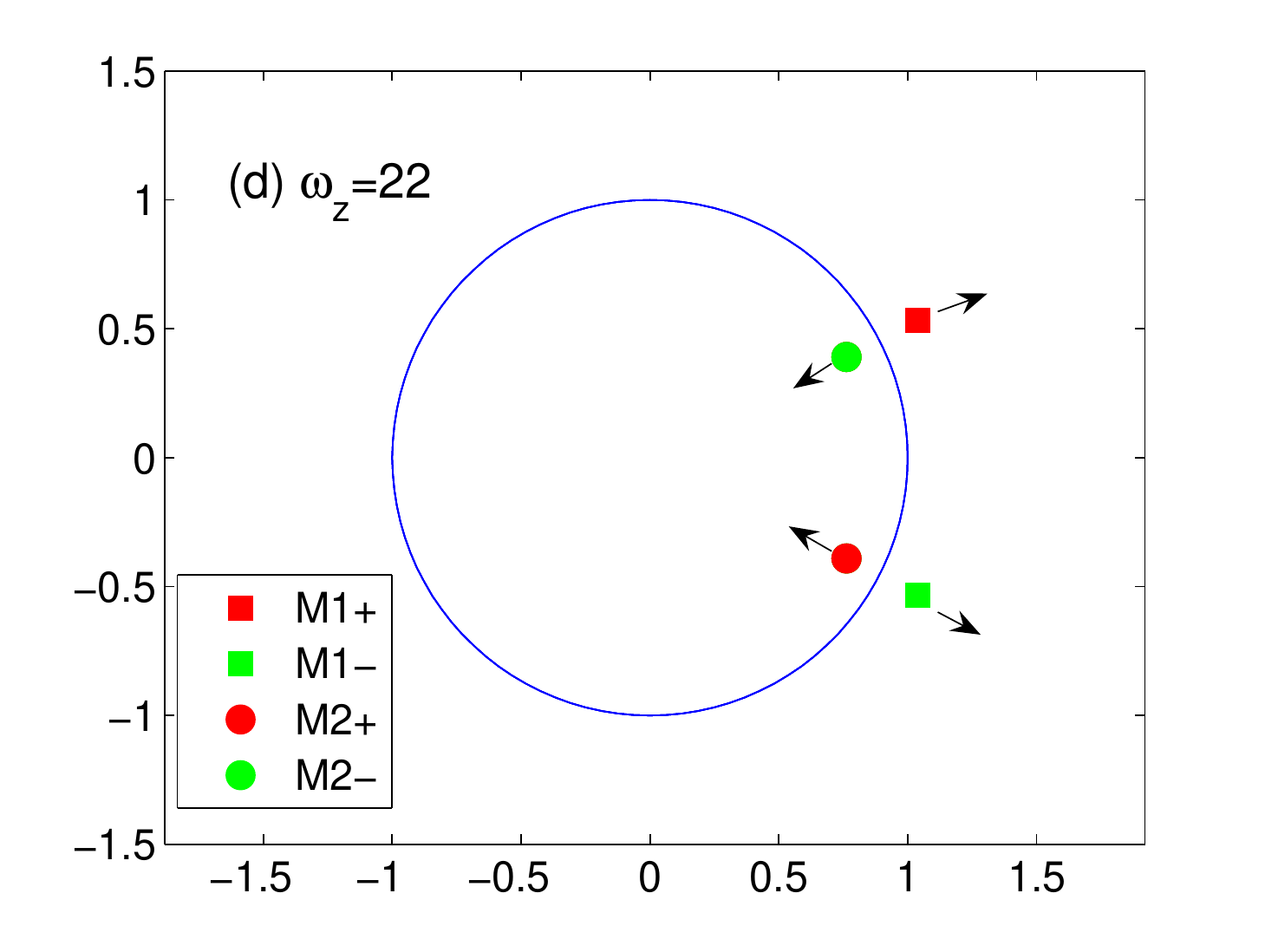}

\caption{Krein collisions when varying $\omega_{z}$ downward at $\omega_{x}=1,\thinspace\omega_{y}=20,\thinspace\mu_{x}=-\mu_{y}/1836$,
$\mu_{y}=E_{0}$ and $E_{0}=320$. Krein collision occurs at $\omega_{z}=22.492$.}
\label{f1}
\end{figure}

To motivate the discussion, we write down the following set of ordinary
differential equations (ODEs) that governs the three-wave instability
in the linear phase \cite{nishikawa1968parametric},
\begin{align}
\frac{d^{2}x}{dt^{2}} & +\omega_{x}^{2}x=\lambda_{x}yz(t),\label{eq:Ne1}\\
\frac{d^{2}y}{dt^{2}} & +\omega_{y}^{2}y=\lambda_{y}xz(t).\label{eq:Ne2}
\end{align}
Here, $x(t)$ and $y(t)$ are the normalized complex amplitudes of
two waves coupled together by a pump wave $z(t)=2\cos(\omega_{z}t)$,
which is given as a known time-dependent function with frequency $\omega_{z}$.
The mode frequencies of the $x-$wave and the $y-$wave are $\omega_{x}$
and $\omega_{y}$ respectively, when they are not coupled by the pump
wave $z(t).$ The normalized amplitude of the pump wave $E_{0}$ is
contained in the complex parameters $\lambda_{x}$ and $\lambda_{y}$,
which measure the strength of the coupling. For the range of parameters
typical to the familiar three-wave interaction between a plasma wave,
an ion-acoustic wave and an electromagnetic pump wave, we have calculated
the instability regions by numerically solving Eqs.\,\eqref{eq:Ne1}
and \eqref{eq:Ne2} for different sets of parameters. The physical
parameters are chosen for a typical hydrogen plasma $\omega_{x}=1$,
$\omega_{y}=20$, $\lambda_{x}/\lambda_{y}=-1/1836$ and $\lambda_{y}=iE_{0}$
(see Figs.\,\ref{f0} and \ref{f1} for the values of normalized
parameters). In Fig.\,\ref{f0}, the instability regions are plotted
in terms of the normalized pump wave frequency $\omega_{z}$ as a
function of the normalized pump strength $E_{0}$. It is discovered
that there are many instability bands, only five of which are shown
in Fig.\,\ref{f0}. There are numerous narrow instability bands below
the lowest band plotted. For a given value of $E_{0}$, the instability
region consists of disconnected intervals. The instability bands can
be viewed as originated from the points on the vertical axis. As $E_{0}$
increases, the instability intervals in terms of $\omega_{z}$ become
larger. The uppermost instability band originates from the position
satisfying condition \eqref{eq:rc}, but the other instability bands
do not. It is interesting that there are two instability bands originated
from $\omega_{z}=\omega_{y}$ and $\omega_{z}=\omega_{x}$ on the
vertical axis.

\begin{figure}
\includegraphics[scale=0.65]{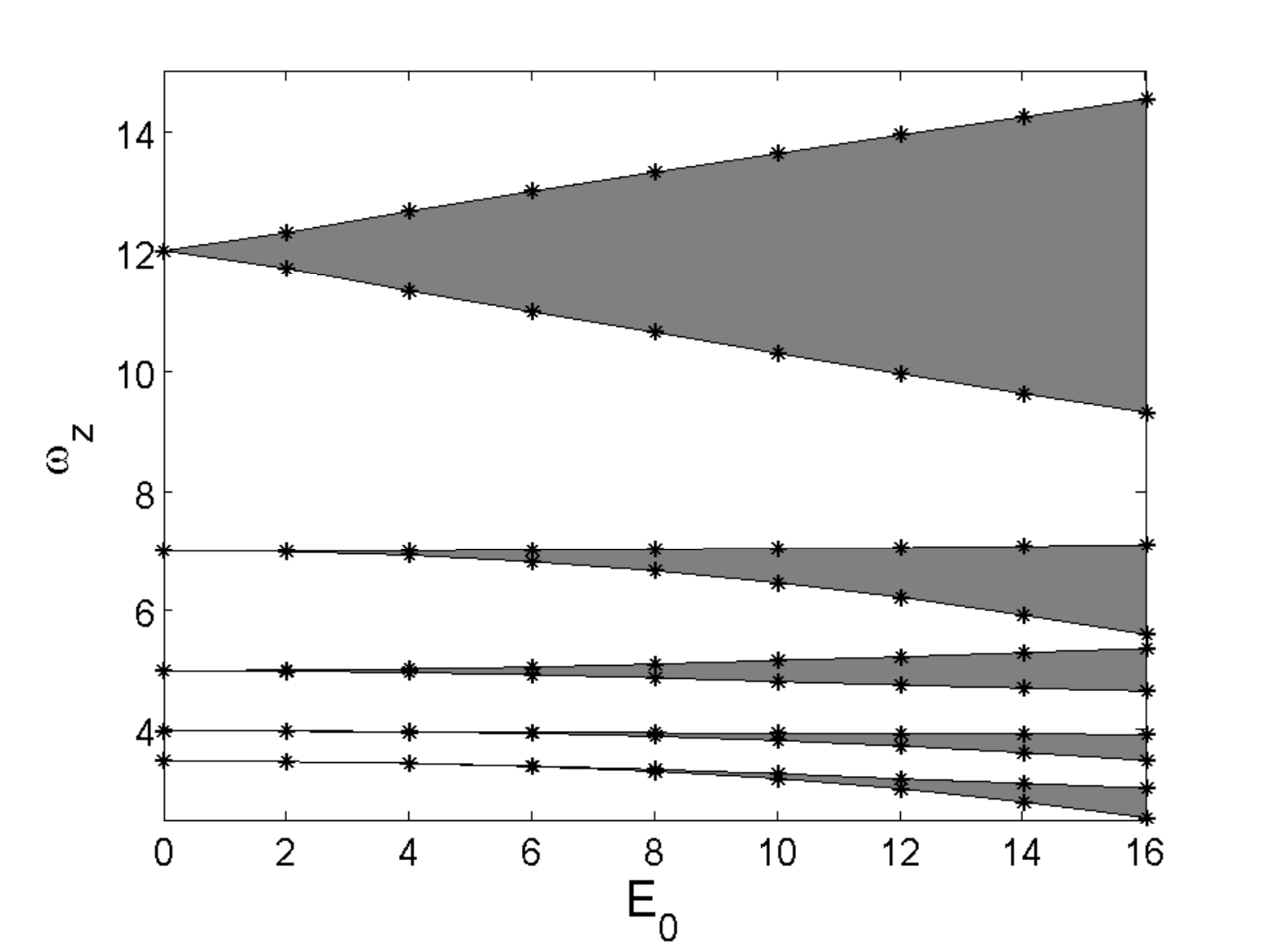}

\caption{Plot of instability regions represented by the normalized frequency
of the pump wave $\omega_{z}$ as a function of the normalized strength
of the pump wave $E_{0}$. Parameters are chosen for the three-wave
interaction with $\omega_{x}=5$, $\omega_{y}=7$, $\lambda_{x}/\lambda_{y}=-1$
and $\lambda_{y}=i\mu_{y}=iE_{0}$. Shaded regions are unstable. Five
instability bands are shown. There exit numerous narrow instability
bands under the lowest band shown. The top three bands originate from
$\omega_{z}=\omega_{x}+\omega_{y}$, $\omega_{z}=\omega_{y}$ and
$\omega_{z}=\omega_{x}$ on the vertical axis. Only the uppermost
instability band originates from the position satisfying condition
\eqref{eq:rc} on the vertical axis. }
\label{f0-1}
\end{figure}

Because the large mass-ratio between protons and electrons, most instability
bands are much lower and narrower than the top two bands. To remove
the nonessential effect due to the large mass-ratio, a different set
of parameters is chosen as $\omega_{x}=5$, $\omega_{y}=7$, $\lambda_{x}/\lambda_{y}=-1$
and $\lambda_{y}=iE_{0}$, which corresponds to an electron-positron
pair plasma with different temperatures for the electrons and positrons.
Many instability bands are found for this case as well. The top five
instability bands are plotted in Fig.\,\ref{f0-1}. We find that
the characters of instability bands are similar to those in Fig.\,\ref{f0}.
Note that the top three bands originate from $\omega_{z}=\omega_{x}+\omega_{y}$,
$\omega_{z}=\omega_{y}$ and $\omega_{z}=\omega_{x}$ on the vertical
axis, as in Fig.\,\ref{f0}. From the two cases plotted Figs.\,\ref{f0}
and \ref{f0-1}, it is evident that the resonant condition \eqref{eq:rc}
cannot be used to characterize the complicated band structures for
the instability. Only when the system parameter approaches the non-interacting
limit, i.e., when $E_{0}\rightarrow0,$ the uppermost instability
band satisfies condition \eqref{eq:rc}.

The existence of other instability bands are unexpected, if not totally
surprising. One may attempt to argue that at $E_{0}\rightarrow0,$
the other bands are described by the generalized resonant condition
\begin{equation}
n_{z}\omega_{z}\approx n_{x1}\omega_{x}-n_{x2}\omega_{x}+n_{y1}\omega_{y}-n_{y2}\omega_{y}.\label{eq:grc}
\end{equation}
In Fig.\,\ref{f0}, the four lower instability bands can be characterized
by $(n_{z},n_{x1},n_{x2},n_{y1},n_{y2})=(1,1,1,1,0)$, $(n_{z},n_{x1},n_{x2},n_{y1},n_{y2})=(1,1,0,1,1)$,
$(n_{z},n_{x1},n_{x2},n_{y1},n_{y2})=(2,1,0,1,1)$ and $(n_{z},n_{x1},n_{x2},n_{y1},n_{y2})=(3,1,0,1,1)$.
In Fig.\,\ref{f0-1}, the four lower instability bands can be characterized
by $(n_{z},n_{x1},n_{x2},n_{y1},n_{y2})=(1,1,1,1,0)$, $(n_{z},n_{x1},n_{x2},n_{y1},n_{y2})=(1,1,0,1,1)$,
$(n_{z},n_{x1},n_{x2},n_{y1},n_{y2})=(3,1,0,1,0)$ and $(n_{z},n_{x1},n_{x2},n_{y1},n_{y2})=(2,1,1,1,0)$.
However, why don't other possible resonances, such as $(n_{z},n_{x1},n_{x2},n_{y1},n_{y2})=(1,0,1,1,0)$
in Fig.\,\ref{f0} or $(n_{z},n_{x1},n_{x2},n_{y1},n_{y2})=(2,1,0,1,0)$
in Fig.\,\ref{f0-1}, appear? We recall that Mathieu's equation has
infinite number of instability bands. For general 2D Hill's equations
\cite{Magnus76}, the phenomena of disappearing unstable intervals
has been noticed and studied \cite{Brown13}. We speculate that there
also exists infinite instability bands for the three-wave interaction
process and that the mechanism of disappearing unstable intervals
is of the same nature with 2D Hill's equations. However, Eqs.\,\eqref{eq:Ne1}
and \eqref{eq:Ne2} are 4D and the corresponding mathematical analysis
can be much more difficult. We are not aware of any previous study
on this topic.

If we discard the resonant condition \eqref{eq:rc} or \eqref{eq:grc}
as the criteria of the onset of the three-wave instability, then what
should be the correct criteria and what is the corresponding physical
mechanism? We will show that the physical mechanism of the three-wave
instability is the resonance between a positive-action mode and a
negative-action mode of the system, and such resonances mark exactly
the instability thresholds. This physical mechanism is a direct consequence
of the Hamiltonian nature of the three-wave interaction process. Mathematically,
however, the Hamiltonian nature is manifested as a non-canonical complex
G-Hamiltonian structure, instead of the familiar real canonical Hamiltonian
structure. The mathematical theory of complex G-Hamiltonian system
was systematically developed by Krein, Gel'fand and Lidskii \cite{Krein1950,Gel1955,KGML1958}.
It is a celebrated result that the G-Hamiltonian system becomes unstable
when and only when two stable eigen-modes of the system with opposite
Krein signatures have the same eigen-frequency, a process known as
Krein collision. It was first discovered in 2016 \cite{zhang2016structure}
that the dynamics of the well known two-stream instability and Jean's
instability are complex G-Hamiltonian in nature, and the instabilities
are results of the Krein collision, which in terms of physics are
found to be resonances between positive- and negative-action modes.
It was also postulated \cite{zhang2016structure} that this is a universal
mechanism for instabilities in Hamiltonian systems with infinite number
of degrees of freedom in plasma physics, accelerator physics and fluid
dynamics. Recently, this is found to be true for the magneto-rotational
instability \cite{kirillov2017singular}. We note that for the special
case of real canonical Hamiltonian system, the Krein collision of
a complex G-Hamiltonian system reduces to the well-known Hopf-Hamilton
bifurcation, which has been identified by Crabtree et al. as the mechanism
of wave-particle instabilities in whistler waves \cite{crabtree2017analysis}.

We would like to point out that the physical mechanism of resonance
between modes with opposite signs of action for the onset of instability
is imposed by the topological and geometric properties of the dynamics
in the spectral space. Resonance is necessary for the onset of instability,
because the eigenvalues of the modes cannot move off the unit circle
without collisions, as mandated by the fact that eigenvalues are symmetric
with respect to the unit circle. This is a topological constraint.
On the other hand, the requirement of opposite signs of action for
the colliding modes is a geometric one.

We start our investigation from Eqs.\,\eqref{eq:Ne1} and \eqref{eq:Ne2}.
Without losing generality, we focus on the three-wave interaction
between a plasma wave, an ion-acoustic wave and an electromagnetic
wave in an unmagnetized plasma \cite{nishikawa1968parametric,nishikawa1968parametric2}.
After normalizing independent variable $t$ and frequencies by the
ion-acoustic frequency $\Omega_{k}$, we obtain the normalized ion-acoustic
frequency $\omega_{x}=1$, the normalized plasma frequency $\omega_{y}=\omega_{e}/\Omega_{k}$
and the normalized frequency of the pump wave $\omega_{z}=\omega/\Omega_{k}$.
The normalized coupling constants are $\lambda_{x}=i\mu_{x}$ and
$\lambda_{y}=i\mu_{y}$ with $\mu_{x}=-\mu_{y}m_{e}/m_{i}$ and $\mu_{y}=E_{0}=ekE/m_{e}\Omega_{k}^{2}$.
Here, $E$ is the amplitude of the pump wave and $E_{0}$ is the normalized
amplitude, and $\mu_{x}$ or $\mu_{y}$ measures the strength of the
pump wave. In this case, the pump wave is assumed to be spatially
uniform (or with a long wavelength), and the plasma wave and the ion
acoustic wave have opposite wave number $k$. In this normalization,
the normalized variable $z(t)=2\cos(\omega_{z}t)$ describes the time-dependent
pump wave. Also, we only consider the case without damping in the
present study.

Equations \eqref{eq:Ne1} and \eqref{eq:Ne2} can be written as a
4-dimensional linear complex dynamical system

\begin{equation}
\mathbf{\dot{\mathbf{x}}}=A(t)\mathbf{x}\thinspace,\label{eq:ls}
\end{equation}
where
\begin{align}
A(t) & =\left(\begin{array}{cccc}
0 & 0 & -\omega_{x}^{2} & i\mu_{x}z(t)\\
0 & 0 & i\mu_{y}z(t) & -\omega_{y}^{2}\\
1 & 0 & 0 & 0\\
0 & 1 & 0 & 0
\end{array}\right)\thinspace,\label{eq:At}\\
\mathbf{x} & =\left(\begin{array}{cccc}
dx/dt & dy/dt & x & y\end{array}\right)^{T}\thinspace.\label{eqzbar-1}
\end{align}
A crucially important property of $A(t)$ is that it is a G-Hamiltonian
matrix, meaning that it can be expressed as
\begin{equation}
A(t)=iG^{-1}S(t)\thinspace,
\end{equation}
for a time-dependent Hermitian matrix $S(t)$ and an invertable constant
Hermitian matrix $G$. This condition is equivalent to that of $A(t)$
satisfying
\begin{equation}
A(t)^{*}G+GA(t)=0\thinspace,\label{eq:GH}
\end{equation}
where $A(t)^{*}$ is the Hermitian conjugate of $A(t).$ The fact
that $A(t)$ is a G-Hamiltonian matrix is the cornerstone of the physical
mechanism of the three-wave instability. It brings out the Hamiltonian
nature of the dynamics and thus determines the stability properties
of system. By analyzing the expression of $A(t),$ we find that
\begin{equation}
G=\left(\begin{array}{cccc}
0 & 0 & -i\mu_{y} & 0\\
0 & 0 & 0 & i\mu_{x}\\
i\mu_{y} & 0 & 0 & 0\\
0 & -i\mu_{x} & 0 & 0
\end{array}\right)\thinspace
\end{equation}
and
\begin{equation}
S(t)=\left(\begin{array}{cccc}
-\mu_{y} & 0 & 0 & 0\\
0 & \mu_{x} & 0 & 0\\
0 & 0 & -\mu_{y}\omega_{x}^{2} & i\mu_{x}\mu_{y}z(t)\\
0 & 0 & -i\mu_{x}\mu_{y}z(t) & \mu_{x}\omega_{y}^{2}
\end{array}\right).
\end{equation}
Equation \eqref{eq:ls} is a special linear G-Hamiltonian system with
the associated energy
\begin{equation}
H(\mathbf{x},t)=-\mathbf{x}^{*}S(t)\mathbf{x}\thinspace,\label{eq:ed}
\end{equation}
where $\mathbf{x}^{*}$ is the conjugate transpose of the vector $\mathbf{x}$.
The definition of general G-Hamiltonian system has been given in Ref.\,\cite{zhang2016structure}.
For $\mathbf{z}\in C^{n}$ and $\bar{\mathbf{z}}\in C^{n}$, the G-Hamiltonian
system is a nature generation of complex Hamiltonian system, and it
reads
\begin{align}
\dot{\mathbf{z}} & =\dfrac{1}{i}G^{-1}\dfrac{\partial H}{\partial\bar{\mathbf{z}}},\label{eq:z-1}\\
\dot{\bar{\mathbf{z}}} & =-\dfrac{1}{i}\bar{G}^{-1}\dfrac{\partial H}{\partial\mathbf{z}},\label{eqzbar-2}
\end{align}
where $G$ is a non-singular Hermitian matrix and $H(\mathbf{z},\bar{\mathbf{z}})$
satisfies the reality condition $H(\mathbf{z},\bar{\mathbf{z}})=\overline{H(\mathbf{z},\bar{\mathbf{z}})}$.
Because of the reality condition, Eq.\eqref{eq:z-1} and Eq.\eqref{eqzbar-2}
are equivalent, and we only need to consider Eq.\eqref{eq:z-1}.

Solutions of Eq.\,\eqref{eq:ls} can be given as $\mathbf{x}(t)=X(t)\mathbf{x}(0)$
for a time evolution matrix, or solution map matrix, $X(t)$. Obviously,
$X(t)$ satisfies
\begin{equation}
\dfrac{dX(t)}{dt}=A(t)X(t)\thinspace,
\end{equation}
and the initial value $X(0)=I$. Since $A(t)$ is a G-Hamiltonian
matrix, the corresponding time evolution matrix $X(t)$ is a G-unitary
matrix, which means that it satisfies
\begin{equation}
X^{*}(t)GX(t)=G\thinspace.\label{eq:G-u}
\end{equation}
This is because
\begin{equation}
\begin{alignedat}{1} & \dfrac{d}{dt}\left[X(t)^{*}GX(t)\right]=\dfrac{d}{dt}X(t)^{*}GX(t)+X(t)^{*}G\dfrac{d}{d}X(t)\\
= & X(t)^{*}A^{*}GX(t)+X^{*}GAX(t)=0\thinspace.
\end{alignedat}
\end{equation}
The eigenvalues of a G-unitary matrix is classified according to their
Krein signatures, which is similar to the definition of the eigenvalues
of a G-Hamiltonian matrix \cite{zhang2016structure}. It is associated
with a bilinear product
\begin{equation}
\left\langle \mathbf{\psi},\phi\right\rangle =\phi^{*}G\psi\thinspace.\label{eq:sd}
\end{equation}
An $r-$fold eigenvalue $\rho$ ($|\rho|=1$) of a G-unitary matrix
with its eigen-subspace $V_{\rho}$ is called the first kind of eigenvalue
if $\left\langle \mathbf{y},\mathbf{y}\right\rangle >0$, for any
$\mathbf{y}\neq0$ in $V_{\rho}$, and the second kind of eigenvalue
if $\left\langle \mathbf{y},\mathbf{y}\right\rangle <0$, for any
$\mathbf{y}\neq0$ in $V_{\rho}$. If there exists a $\mathbf{y}\in V_{\rho}$
such that $\left\langle \mathbf{y},\mathbf{y}\right\rangle =0$, then
$\rho$ is called an eigenvalue of mixed kind \cite{KGML1958}. The
first kind and the second kind are also called definite, and the mixed
kind is also called indefinite.

In the present case, $A(t)$ is periodic with period $T=2\pi/\omega_{z}$.
So is the solution map matrix $X(t)$, and thus the dynamic properties
of the system are given by the one-period map $X(T)$. Specifically,
eigenvalues of $X(T)$ decide the stability property of the system.
The relevant mathematical theory has been systematically developed
by Krein, Gel'fand and Lidskii \cite{Krein1950,Gel1955,KGML1958},
which is listed as follows.
\begin{enumerate}
\item \emph{The eigenvalues of a G-unitary matrix are symmetric with respect
to the unit circle.}
\item \emph{The number of each kind of eigenvalue is determined by the Hermitian
matrix $G$. Let p be the number of positive eigenvalues and q be
the number of negative eigenvalues of the matrix $G$, then any G-unitary
matrix has p eigenvalues of first kind and q eigenvalues of second
kind (counting multiplicity).}
\item \emph{(Krein-Gel'fand-Lidskii theorem) The G-Hamiltonian system is
strongly stable if and only if all of the eigenvalues of the one-period
map matrix $X(T)$ lie on the unit circle and are definite. Here,
strongly stable means that the system is stable in an open neighborhood
of the parameter space constrained by the G-Hamiltonian structure. }
\end{enumerate}
From the above mathematical results, we know that the eigenvalues
of one-period evolution map $X(T)$ of a G-Hamiltonian system are
symmetric about the unit circle. Moreover, because the eigenvalues
of $G$ are $\pm\mu_{x},\pm\mu_{y}$, two of which are positive and
the other two are negative, the G-unitary matrix $X(T)$ has two eigenvalues
of the first kind and two eigenvalues of the second kind. When two
eigenvalues of $X(T)$ with different Krein signatures collide on
the unit circle, the G-Hamiltonian system Eq.\,\eqref{eq:ls} is
destabilized. This process is called Krein collision. We now show
its physical meaning. For a G-unitary matrix $X(T)$ whose eigenvalues
are distinct and are all on the unit circle, it can be written as
\begin{equation}
X(T)=YDiag(e^{i\lambda_{1}},\cdots,e^{i\lambda_{4}})Y^{-1}\thinspace,
\end{equation}
where $\lambda_{i}\in[0,2\pi]$ and $Diag(e^{i\lambda_{1}},\cdots,e^{i\lambda_{4}})$
is a diagonal matrix. We define
\begin{equation}
\ln X(T)=YDiag(i\lambda_{1},\cdots i\lambda_{4})Y^{-1}\thinspace.\label{eq:lnd}
\end{equation}
Because $X(T)$ is a G-unitary matrix, we have
\begin{equation}
G^{-1}\left[\ln X(T)\right]^{*}G=\ln\left[G^{-1}X(T)^{*}G\right]=\ln\left[X(T)^{-1}\right]=-\ln X(T)\thinspace.
\end{equation}
Thus $\ln X(T)/T$ is a G-Hamiltonian matrix and there is a Hermitian
matrix $\hat{S}$, such that $\ln X(T)/T=iG^{-1}\hat{S}\thinspace.$
For an eigenvalue $\rho=\exp(i\lambda)$ of $X(T)$ on the unit circle
with an eigenvector $\mathbf{y}$,
\begin{align}
\mathbf{y}^{*}G\dfrac{\ln X(T)}{T}\mathbf{y} & =i\mathbf{y}^{*}\hat{S}\mathbf{y}\thinspace.\label{eq:a1}
\end{align}
Because of Eqs.\,\eqref{eq:ed} and \eqref{eq:sd}, we obtain
\begin{equation}
<\mathbf{y},\mathbf{y}>=-\dfrac{T\hat{H}(\mathbf{y})}{\lambda}\thinspace.
\end{equation}
Here, $T\hat{H}(y)$ is the action over one period and $\lambda$
is the argument of the eigenvalue $\rho$. It is clear that in a strongly
stable system, the physical meaning of its Krein signature is the
opposite sign of the action over one period. Therefore, the physical
mechanism of the\emph{ }Krein-Gel'fand-Lidskii theorem is that the
system becomes unstable when and only when a positive-action mode
resonates with a negative-action mode.

\begin{figure}
\includegraphics[scale=0.6]{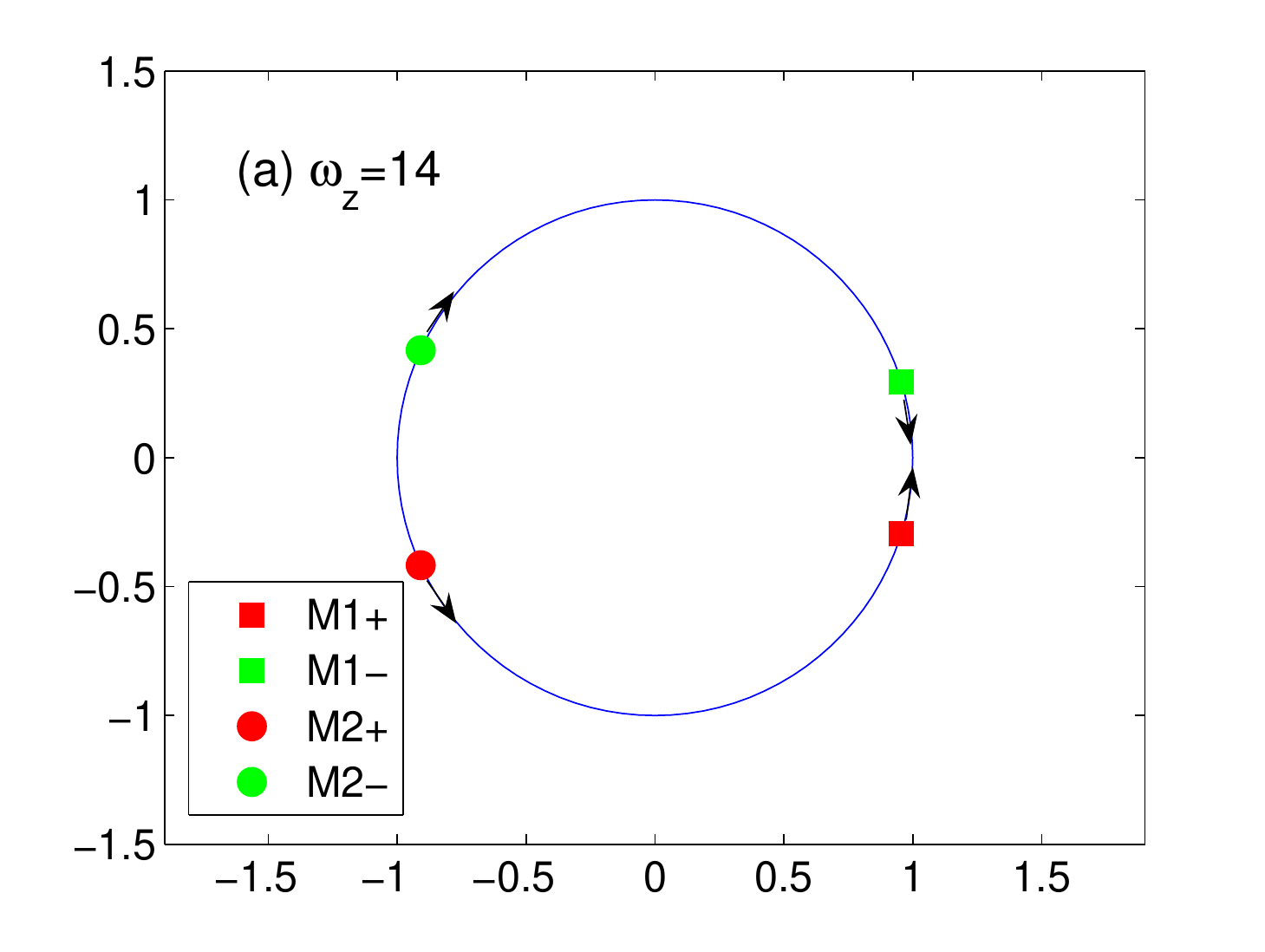}\includegraphics[scale=0.6]{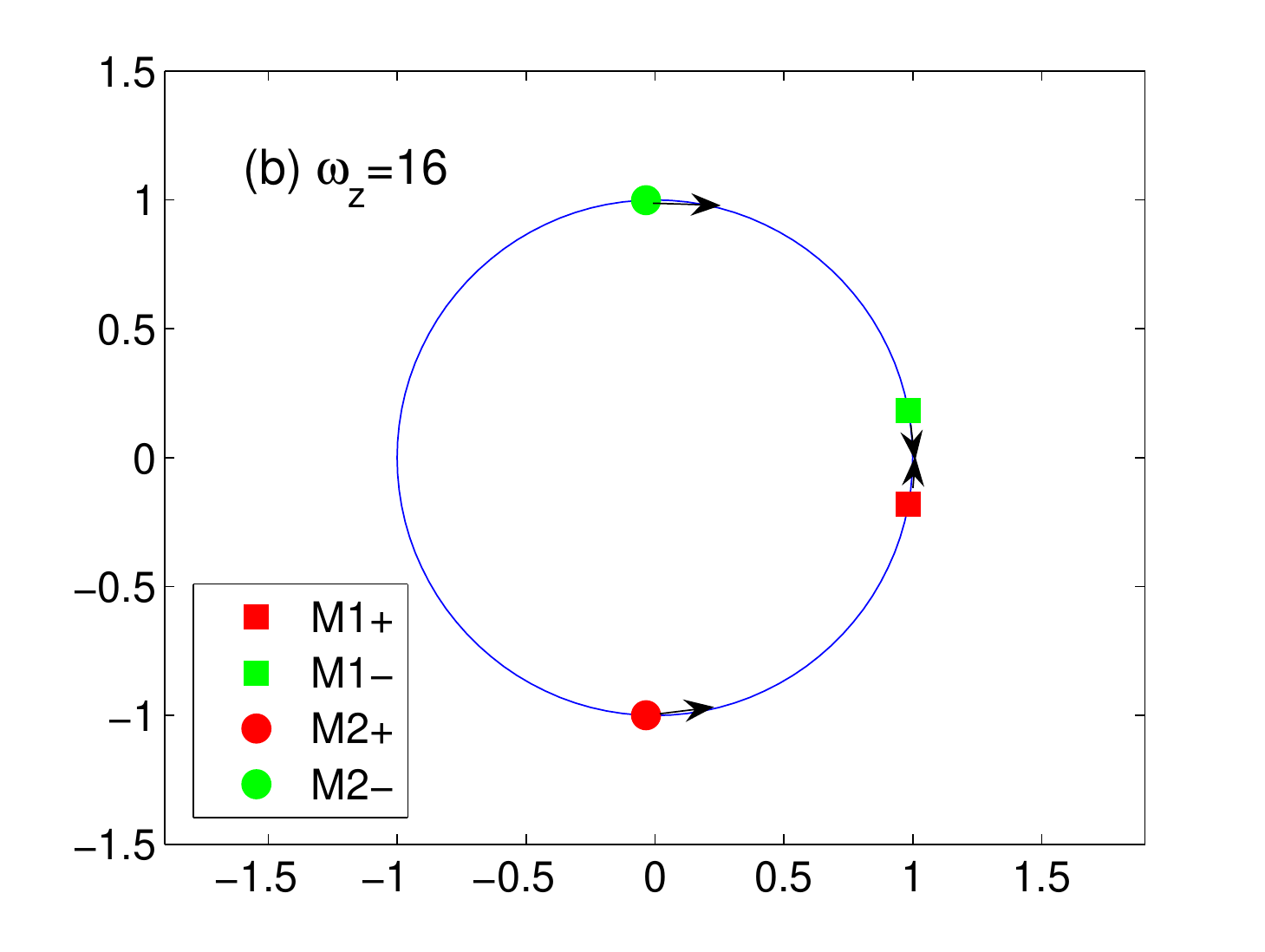}

\includegraphics[scale=0.6]{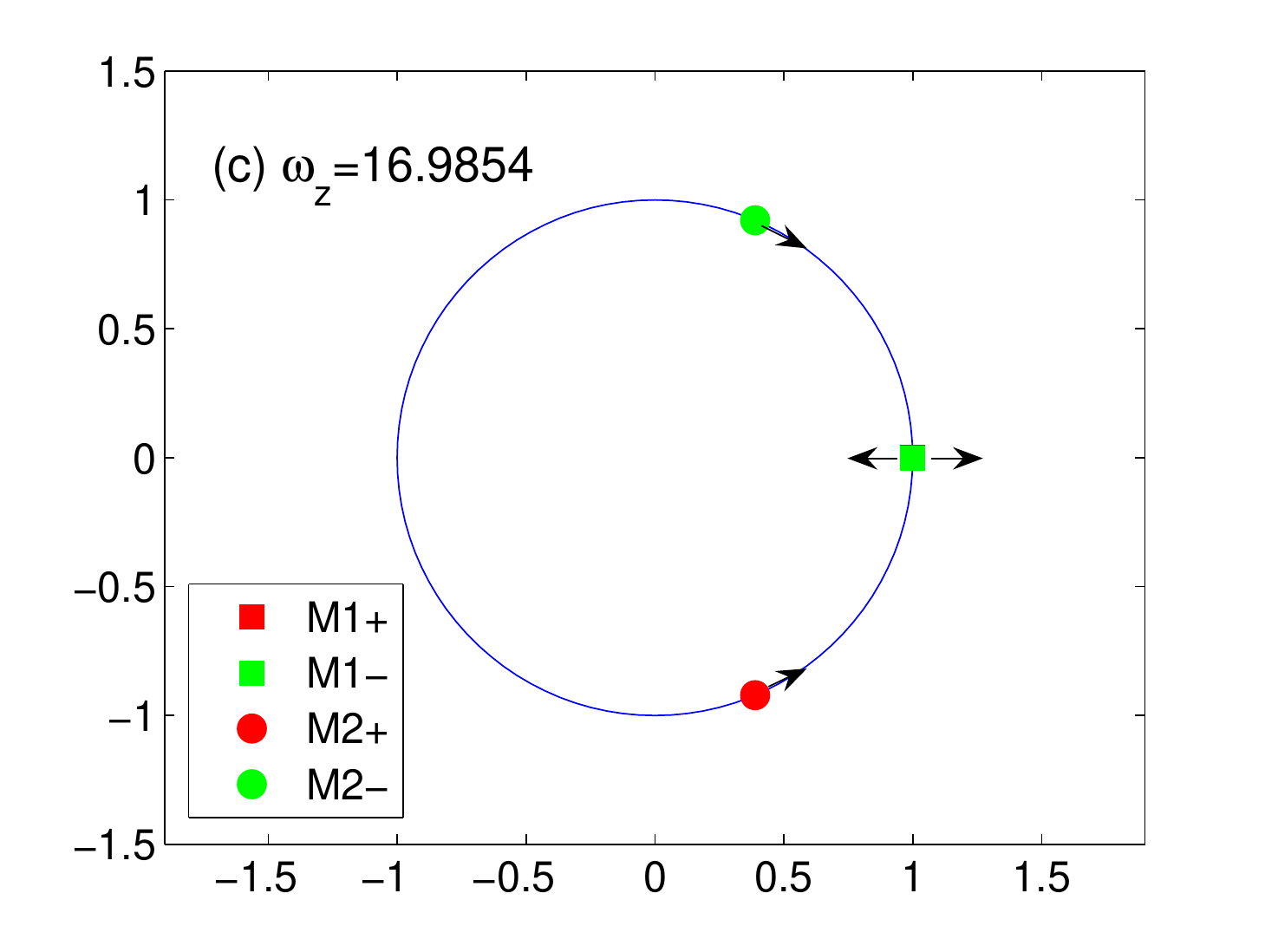}\includegraphics[scale=0.6]{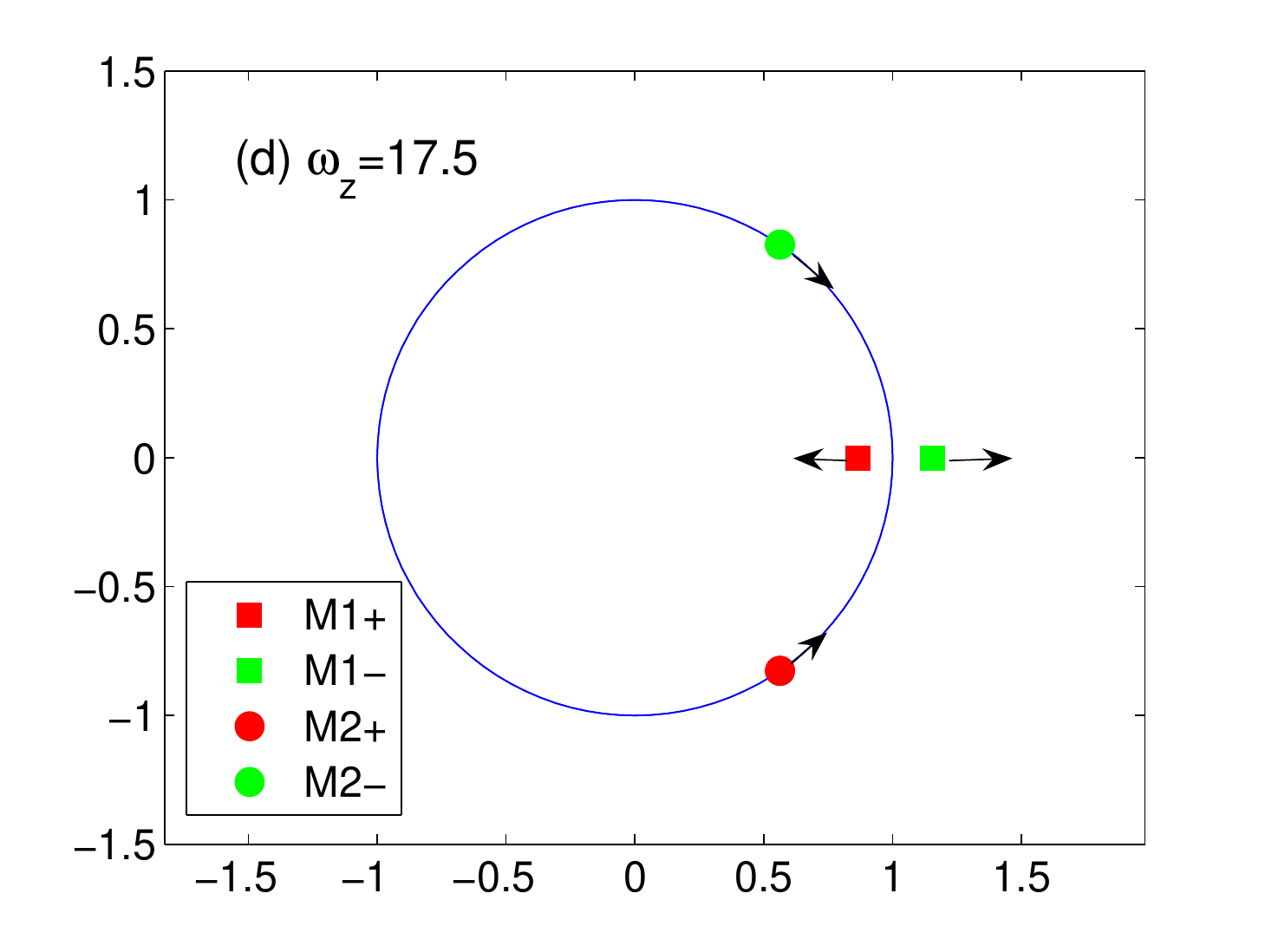}

\caption{Krein collisions when varying $\omega_{z}$ upward at $\omega_{x}=1,\thinspace\omega_{y}=20,\thinspace\mu_{x}=-\mu_{y}/1836$,
$\mu_{y}=E_{0}$ and $E_{0}=320$. Krein collision occurs at $\omega_{z}=16.9854$.}
\label{f2}
\end{figure}

We now demonstrate the aforementioned physical mechanism using numerically
calculated examples for a typical hydrogen plasma. The normalized
plasma frequency is chosen to be $\omega_{y}=20,$ and the normalized
pump amplitude $\mu_{y}=E_{0}$ varies from $0$ to $400.$ For each
$E_{0}$, the system is solved numerically for different values of
$\omega_{z}$. The one-period evolution map $X(T)$ and its eigenvalues
are calculated numerically. The instability regions in terms of frequency
$\omega_{z}$ are plotted as functions of $E_{0}$ in Fig.\,\ref{f0}.
As discussed above, it shows clearly that resonant condition \eqref{eq:rc}
cannot be used as the criteria for the onset of the three-wave instability
in general. Instead, the instability thresholds are exactly the locations
where a positive-action mode resonates a negative-action mode, which
is shown in Figs.\,\ref{f1} and \ref{f2}.

For the case of $E_{0}=320$, five instability intervals are shown
in Fig.\,\ref{f0}. First, let's observe how the system behaves as
$\omega_{z}$ traverses the uppermost instability threshold from the
top. Shown in Fig.\,\ref{f1} are the eigenmodes numerically calculated.
As expected, the eigenvalues of the matrix $X(T)$ are symmetric about
unit circle and real axis. There are two sets, and each set contains
two eigenvalues symmetric about real axis. When $\omega_{z}=24$,
shown in Fig.\,\ref{f1}(a), the eigenmodes are all on the unit circle
and distinct, where $M_{1+}$ and $M_{2+}$ (marked by red) are eigenmodes
with negative action, and $M_{1-}$ and $M_{2-}$ (marked by green)
are eigenmodes with positive action. As $\omega_{z}$ decreases, the
eigenmodes $M_{1+}$ and $M_{2+}$ rotate clockwise on the unit circle,
meanwhile $M_{1-}$ and $M_{2-}$ rotate counterclockwise on the unit
circle as shown in Fig.\,\ref{f1}(b). Decreasing $\omega_{z}$ to
$22.492$, we find in Fig.\,\ref{f1}(c) that eigenmodes $M_{1+}$
and $M_{2-}$ collide on the unit circle, and simultaneously $M_{1-}$
and $M_{2+}$ collide on the unit circle. This marks the uppermost
threshold of the instability. Then, when $\omega_{z}=22$, all eigenmodes
move off the unit circle with $M_{1+}$ and $M_{1-}$ outside and
$M_{2+}$ and $M_{2-}$ inside, as shown in Fig.\,\ref{f1}(d). The
system is now unstable, and $M_{1}$ and $M_{2}$ are the unstable
modes.

Similarly, we can move $\omega_{z}$ upwards traversing the lower
threshold of the second instability band. When $\omega_{z}=14$, shown
in Fig.\,\ref{f2}(a), the system is stable with four eigenmodes
on the unit circle. As $\omega_{z}$ increases, $M_{1+}$ and $M_{1-}$
move towards each other as shown in Fig.\,\ref{f2}(b), and they
collide when $\omega_{z}=16.9854$, which marks the lowermost threshold
of the instability in Fig.\,\ref{f2}(c). Increasing $\omega_{z}$
to $17.5$, the $M_{1+}$ and $M_{1-}$ modes become unstable in Fig.\,\ref{f2}(d).
In the process, the $M_{2+}$ and $M_{2-}$ modes stay on the unit
circle all the time, which is different from the case displayed in
Fig.\,\ref{f1}.

In conclusion, we have shown that for the dynamics of three-wave interaction,
the familiar resonant condition $\omega_{z}\approx\omega_{x}+\omega_{y}$
is not the criteria for instability. The physical mechanism of the
instability is the resonance between a positive-action mode with a
negative-action mode, and this condition exactly marks the instability
threshold. This mechanism is imposed by the topology and geometry
of the spectral space determined by the complex G-Hamiltonian structure
of the dynamics, which is a manifest of the infinite dimensional Hamiltonian
structure in the wave-number space. Guided by this new theory, the
additional instability bands for the three-wave interaction previously
unknown are discovered. They originate from points on the vertical
axis. We suspect that the mechanism of resonance between a positive-
and negative-action modes may be crucial in predicting which point
instability bands originate from. Study on this and other topics will
be reported in the future.

This research is supported by the National Natural Science Foundation
of China (NSFC-11505186, 11575185, 11575186), ITER-China Program (2015GB111003,
2014GB124005), the Geo-Algorithmic Plasma Simulator (GAPS) Project,
and the U.S. Department of Energy (DE-AC02-09CH11466).

%

\end{document}